\documentclass[12pt]{iopart}
\usepackage{iopams}
\usepackage{graphicx}
\usepackage{bm}
\def\la{{\langle}}
\def\ra{{\rangle}}

\newcommand{\beq}{\begin{equation}}
\newcommand{\eeq}{\end{equation}}
\newcommand{\beqa}{\begin{eqnarray}}
\newcommand{\eeqa}{\end{eqnarray}}
\newcommand{\cT}{{\cal{T}}}
\newcommand{\beqas}{\begin{eqnarray*}}
\newcommand{\eeqas}{\end{eqnarray*}}

\newcommand {\fcos} [1] {\cos \left( #1 \right)}
\newcommand {\fsin} [1] {\sin \left( #1 \right)}
\newcommand {\fabs}[1] {\left| #1 \right|}
\newcommand {\fabsq}[1] {\left| #1 \right|^2}

\begin{document}
\title{Optimally robust shortcuts to population inversion in 
two-level quantum systems}

\author{A. Ruschhaupt$^{1,2}$, Xi Chen$^{3,4}$, D. Alonso$^5$ and J. G. Muga$^{3,4}$}
\address{$^{1}$ Institut f\"ur Theoretische Physik, Leibniz Universit\"at Hannover, Appelstr. 2, 30167 Hannover, Germany}
\address{$^{2}$ Department of Physics, University College Cork, Ireland}
\address{$^3$ Departamento de Qu\'{\i}mica-F\'{\i}sica, UPV-EHU, Apdo 644, 48080 Bilbao, Spain}
\address{$^4$ Department of Physics, Shanghai University, 200444 Shanghai, China}
\address{$^5$ Departamento de F\'\i sica Fundamental y Experimental,
Electronica y Sistemas and IUdEA, Universidad de La Laguna, 38203 La Laguna, Spain}

\begin{abstract}
We examine the stability versus different types of perturbations of recently proposed shortcuts-to-adiabaticity to speed up
the population inversion of a two-level quantum system. 
We find optimally robust processes using invariant based engineering of the Hamiltonian.  
Amplitude noise
and systematic errors require different optimal protocols.    
\end{abstract}
%
%
%
%
%
%
%
%
%
%
\section{Introduction}
Manipulating the state of a quantum system with time-dependent interacting
fields is a fundamental operation in atomic and molecular physics, with
applications such as laser-controlled chemical reactions, metrology,
interferometry, nuclear magnetic resonance (NMR), or quantum information
processing \cite{Allen,Vitanov-Rev,Bergmann,Kral}.
For two-level systems
there are several approaches proposed to attain a complete population
transfer, for example, $\pi$ pulses, composite pulses,
adiabatic passage and its variants. In general, the $\pi$ pulses may be fast  but highly sensitive to 
variations in the pulse area, and to inhomogeneities in the sample \cite{Allen}.
Used first in nuclear magnetic resonance \cite{Levitt},
composite pulses provide an alternative to the single $\pi$-pulse, 
with some successful applications \cite{Collin,Torosov}, 
but still need an accurate control of pulse phase and intensity.
A robust option is in principle adiabatic (slow) passage, which is 
however prone to decoherence because of the effect of noise  
over the long times required. A compromise is to use  
speeded-up ``shortcuts to adiabaticity'', which may be broadly defined as the   
processes that lead to the same final populations than the adiabatic
approach in a shorter time. 

Several methods to find  
shortcuts to adiabaticity  have
been put forward \cite{Rice,Berry09,Chen10b,Masuda,JPB10,Chen11,Sara,Oliver,Sara11,Fashi,Lacour} 
for two- and three-level atomic systems.
The transitionless or counter-diabatic control protocols, proposed by Demirplak, Rice \cite{Rice} and Berry \cite{Berry09} start from a reference time-dependent 
Hamiltonian $H_0$ and 
provide an extra interaction that cancels the diabatic couplings. This 
results in an exact following of the adiabatic dynamics of the reference Hamiltonian,
in principle in an arbitrarily short time. They have been applied, for example, to speed up the RAP for an Allen-Eberly scheme \cite{Chen10b}. Modified by a unitary transformation \cite{Sara11}, 
the transitionless quantum driving has been experimentally implemented 
for a two-level system realized by Bose-Einstein condensates in optical lattices \cite{Oliver}.

Another shortcut technique is to inverse engineer the Hamiltonian  
using Lewis-Riesenfeld invariants \cite{LR}, 
as in 
\cite{bec,Chen,energy,Li,Nice,Nice2,Nice3,transport,transport2,opttransport,Wu,Adol,Onofrio}. The invariant-based method has been applied to accelerate the adiabatic processes for trap expansion or compressions \cite{bec,Chen,energy,Li,Nice,Nice2,Nice3} and atomic transport \cite{transport,transport2,opttransport}. It has also been combined  
with optimal control theory \cite{Li,opttransport}, and proposed for other applications \cite{Wu,Adol,Onofrio,Sara11}.
Counterdiabatic and invariant-based engineering can in fact be shown to be potentially 
equivalent methods by properly adjusting the reference Hamiltonian \cite{Chen11}. In standard applications though, $H_0$ is set according to some predetermined, standard protocol
(for example Landau-Zener, Allen-Eberly, or finite-time schemes),  and the
formulation and results of the two methods are generally quite different, so  they may be considered in practice separate approaches.  

A key element to choose among the fast protocols is their stability or robustness 
versus different perturbations. We will compare the results with ordinary
(flat) $\pi$ pulses and explore
the stability of the transitionless approach with respect to parameter
variations for a finite-time sinusoidal protocol for $H_0$. 

The main aim of this paper is to find optimal protocols with respect to amplitude noise of the interaction and with respect to systematic errors. The optimality will be determined by minimizing properly defined sensitivities. It turns out that the perturbations due to noise and systematic errors require different optimal protocols,  and we shall use invariant-based inverse engineering to find them. 

The rest of the paper is organized as follows.
In the following section we shall review the transitionless-based shortcuts
protocol and the invariant-based one. 
In \sref{formalism}, the general formalism to model amplitude-noise error and
systematic error will be presented.
The special case of solely amplitude-noise error will be examined in
\sref{noise_error} where the noise sensitivity of the different protocols will
be studied and the most stable protocol will be derived.
In \sref{systematic_error}, the special case of solely systematic error
will be studied and the most stable protocol will be derived.
The general case of amplitude-noise as well as systematic noise will be for
the different protocols will be numerical studied in \sref{noise_systematic}.

%
%
%
%
\section{Shortcuts to adiabatic passage for a two-level quantum system}
\label{STA}
We assume a two-level system with a Hamiltonian of the form
\beqa
H_0 (t)= \frac{\hbar}{2} \left(\begin{array}{cc} -\Delta(t) & \Omega_{R}(t) -i\Omega_I(t)
\\
\Omega_{R}(t)+i\Omega_I(t) &  \Delta(t)
\end{array}\right).
\label{H0}
\eeqa
For example, in quantum optics such a Hamiltonian describes the semiclassical 
coupling of two atomic levels with a laser in a laser-adapted interaction picture.
In that setting $\Omega(t) = \Omega_R(t) + i \Omega_I(t)$ would be the complex Rabi
frequency (where $\Omega_R$ and $\Omega_I$ and the real and imaginary
parts) and $\Delta$ would be the time-dependent detuning between laser and transition frequencies. We find it convenient to keep the language of the atom-laser interaction
hereafter noting that in other two-level systems $\Omega(t)$ and $\Delta(t)$ will 
correspond to different physical quantities and that instead of ``atom'' 
one may refer, for example, to a spin-$1/2$, or to a Bose-Einstein condensate on an 
accelerated optical lattice \cite{Oliver}.   
      
Initially at time $t=0$, the atom is in the ground state. Often the goal is to
achieve a perfect population inversion such that at a time $t=T$ the atom
should be in the excited state. The
time $T$ should be as small as possible but also the scheme or
protocol to achieve this population inversion should be as stable as possible
concerning errors. In the following subsections we will review different schemes to
achieve a population inversion before we discuss different types of
possible error sources in the next section. 
\subsection{$\pi$ pulse}
A simple scheme to achieve population inversion is a $\pi$ pulse. In this
case the laser is on resonance, i.e. the detuning is zero $\Delta(t)=0$ for
all $t$. If the Rabi frequency is chosen like $\Omega(t) = \fabs{\Omega (t)}
e^{i\alpha}$, with a time-independent $\alpha$, and such that
\beqa
\int_0^T dt\, \fabs{\Omega(t)} = \pi, 
\eeqa
the population is inverted at time $T$.
A simple example is the ``flat'' $\pi$ pulse with $\Omega(t) = \frac{\pi t}{T}
e^{i\alpha}$.

\subsection{Adiabatic schemes and transitionless shortcuts to
adiabaticity}
The population inversion may also be achieved by an adiabatic scheme. Let
the instantaneous eigenstates of the Hamiltonian $H_0$ be $|n(t)\ra$, with $n=0,1$. The
adiabatic theorem tells us that if we start in an eigenstate at $t=0$,
i.e. $|\psi(0)\ra = |n(0)\ra$ and if
we vary the Hamiltonian infinitesimally slowly, then the system will stay in the
corresponding instantaneous eigenstate for all times, up to a phase factor,
i.e. $|\psi(t)\ra \approx e^{i \kappa_n(t)}|n(t)\ra$. If the 
eigenstate corresponds initially to the ground state and at $t=T$ to the excited
state (up to a phase) then we would achieve a perfect population inversion as
$T\to\infty$.

Demirplak and Rice \cite{Rice} and independently Berry \cite{Berry09} proposed
a modification of the Hamiltonian such that the state would exactly follow the instantaneous
eigenstate of the Hamiltonian $H_0$ for an arbitrary duration $T$.  
If the desired time evolution operator is $U = \sum e^{i
\kappa_{n} (t)} |n(t) \rangle \langle n(0)|$, the
corresponding Hamiltonian leading to this time evolution is $H_{0a} (t) = i\hbar
(\partial_t U) U^+$. We may write $H_{0a} = H_0 + H_a$, where $H_a = i\hbar  \sum_n |
\partial_t n(t) \rangle \langle n(t) |$. 
is the ``counter-diabatic'' (CD) term that guarantees that the system will follow the
instantaneous eigenstates of $H_0$ without transitions even for a small $T$.
This method is thus termed counter-diabatic approach or transitionless-tracking algorithm. 

For the two-level
system with Hamiltonian $H_0$ and $\Omega_I = 0$ this additional Hamiltonian
takes the the form
\beqa
H_a (t)
= \frac{\hbar}{2} \left(\begin{array}{cc} 0 & -i \Omega_{a}(t)
\\
i \Omega_{a}(t) & 0
\end{array}\right),
\label{h1s}
\eeqa
with $\Omega_a \equiv [\Omega_R \dot{\Delta} - \dot{\Omega}_R \Delta]/\Omega^2$.
The total Hamiltonian is therefore \cite{Chen10b}
\beqa
H_{0a} (t)= \frac{\hbar}{2} \left(\begin{array}{cc} -\Delta & \Omega_{R}- i  \Omega_a
\\
\Omega_{R}+ i  \Omega_a &  \Delta
\end{array}\right), 
\eeqa
which we will call transitionless shortcut protocol in the following.

\subsection{Inverse engineering of invariant-based shortcuts}
Shortcuts to adiabaticity can be also found making 
explicit use of Lewis-Riesenfeld
invariants. 
For the general Hamiltonian $H_0$ in (\ref{H0}), 
a dynamical invariant of the corresponding Schr\"odinger
equation (this is a Hermitian Operator $I(t)$ fulfilling $\frac{\partial}{\partial t}I + \frac{i}{\hbar}[H_0,I] = 0$, so that its expectation values remain constant) is given by
\beqa
\label{I}
I (t)= \frac{\hbar}{2} \mu \left(\begin{array}{cc} \fcos{\Theta(t)} & \fsin{\Theta(t)} e^{- i \alpha(t)}
\\ \fsin{\Theta(t)} e^{i \alpha(t)} &  -\fcos{\Theta(t)}
\end{array}\right),
\eeqa
where $\mu$ is an arbitrary constant with units of frequency to keep $I(t)$ with dimensions of energy, and the functions $\Theta(t)$ and $\alpha(t)$
satisfy the differential equations
\begin{eqnarray}
\begin{array}{l}
\dot\theta = \Omega_I \cos \alpha - \Omega_R \sin \alpha,
\\
\dot\alpha =  -\Delta(t) - \cot\Theta\left(\cos\alpha\, \Omega_R +
\sin\alpha\,\Omega_I\right).
\end{array}
\label{schrpure}
\end{eqnarray}
The eigenvectors of the invariant are
\beqa
\fl
|\phi_+(t)\ra = \left(\begin{array}{c} \fcos{\Theta/2} e^{-i\alpha/2}\\
\fsin{\Theta/2}e^{i\alpha/2}
\end{array}\right), \quad
|\phi_-(t)\ra = \left(\begin{array}{c} \fsin{\Theta/2} e^{-i\alpha/2}\\
-\fcos{\Theta/2} e^{i\alpha/2}
\end{array}\right), 
\eeqa
with the eigenvalues $\pm \frac{\hbar}{2} \mu$.
A general solution $|\Psi(t)\ra$ of the Schr\"odinger equation can be written as a linear
combination 
$|\Psi(t)\ra = c_+ e^{i\kappa_+(t)} |\phi_+(t)\ra + c_- e^{i\kappa_-(t)} |\phi_-(t)\ra$,
where $c_\pm$ are complex, constant coefficients, and 
$\kappa_\pm$ are the Lewis-Riesenfeld phases \cite{LR}
\begin{eqnarray*}
\dot\kappa_+ &=& \frac{1}{\hbar} \Big\langle
\phi_+\Big|i\hbar\frac{\partial}{\partial t} - H_0\Big| \phi_+\Big\rangle,\\
\dot\kappa_- &=& \frac{1}{\hbar} \Big\langle
\phi_-\Big|i\hbar\frac{\partial}{\partial t} - H_0\Big| \phi_-\Big\rangle.
\end{eqnarray*} 
In particular we may construct the solution 
\begin{eqnarray}
|\psi(t)\ra = |\phi_+(t)\ra e^{-i \gamma(t)/2} = \left(\begin{array}{c} \cos(\Theta/2) e^{-i\alpha/2}\\
\sin(\Theta/2)e^{i\alpha/2}
\end{array}\right) e^{-i \gamma/2}
\label{purestate}
\end{eqnarray}
and the orthogonal solution (for all times $\la
\psi_\perp|\psi\ra = 0$)
\begin{eqnarray}
|\psi_\perp(t)\ra = |\phi_-(t)\ra e^{i \gamma(t)/2} = \left(\begin{array}{c} \sin(\Theta/2)e^{-i\alpha/2}\\
-\cos(\Theta(t)/2) e^{i\alpha/2}
\end{array}\right) e^{i \gamma/2}.
\label{psip}
\end{eqnarray}
where 
\begin{eqnarray*}
\gamma = - 2 \kappa_+ = 2 \kappa_-.
\end{eqnarray*}
Finally, we get
\begin{eqnarray}
\dot\gamma &=& \frac{1}{\sin\Theta} \left(\cos\alpha\,\Omega_R +
\sin\alpha\,\Omega_I\right).
\label{dotgamma}
\end{eqnarray}
Equivalently we may design a solution of the Schr\"odinger equation
$|\psi(t)\ra$ 
with the parameterization
of a pure state given in (\ref{purestate}). (Note that  
$|\psi(t)\ra\la\psi(t)|$ is a dynamical invariant.)
By putting this ansatz into the Schr\"odinger equation, we get
immediately (\ref{schrpure}) and (\ref{dotgamma}). 
A solution which is orthogonal to (\ref{purestate}), i.e. $\la
\psi_\perp|\psi\ra = 0$ for all times, is then directly given by (\ref{psip}).

The next step to find invariant-based shortcuts is to inverse
engineer the Hamiltonian.
For achieving a population inversion, the boundary values should be
$\Theta(0)=0$ and $\Theta(T)=\pi$, so 
\begin{eqnarray*}
|\psi(0)\ra = \left(\begin{array}{c} e^{-i\alpha(0)/2}\\
0 \end{array}\right) e^{-i \gamma(0)/2}\, ,\quad
|\psi(T)\ra = \left(\begin{array}{c} 0,
\\
e^{i\alpha(T)/2} \end{array}\right) e^{-i \gamma(T)/2}.
\end{eqnarray*}
Assume that $\Theta(t)$ and $\alpha(t)$ are given. Then
we get for the Hamiltonian corresponding to this solution by inverting
(\ref{schrpure}). This leads to
\begin{eqnarray}
\Omega_R &=& \cos\alpha\sin\Theta \;\dot\gamma -
\sin\alpha\;\dot\Theta\label{pot_R},
\\
\Omega_I &=& \sin\alpha\sin\Theta\;\dot\gamma +
\cos\alpha\;\dot\Theta\label{pot_I},
\\
\Delta &=& -\cos\Theta \;\dot\gamma - \dot\alpha.
\label{pot_D}
\end{eqnarray}
By implementing these functions exactly the population would be 
inverted in the unperturbed, error-free case.
Note that invariant-based shortcuts and
transitionless shortcuts may be formally related, see \cite{Chen11}.

%
%
%
%
%
%
\section{General formalism for systematic and amplitude-noise errors}
\label{formalism}
We shall now consider systematic errors as well as noise-related
errors. Let the ideal, unperturbed Hamiltonian be $H_0$. 
For systematic errors, the actual, experimentally implemented 
Hamiltonian is $H_{01}=H_0
+ \beta H_1$, but the evolution of the pure quantum state is still
described by the Schr\"odinger equation with the perturbed Hamiltonian $H_{01}$. 
Sometimes systematic errors cannot be avoided, for example if different atoms
at different positions are subjected to slightly different fields, due to, 
for example, the Gaussian shape of the
laser inducing different Rabi frequencies. 
It is
thus desirable to have protocols which are very stable with respect to  
perturbed Hamiltonian functions. 

The second type of error is a stochastic one, i.e. the Hamiltonian $H_{01}$ is
perturbed by some stochastic part $\lambda H_2$ describing amplitude noise.
A stochastic Schr\"odinger equation (in the Stratonovich sense) is then
\begin{eqnarray*}
i\hbar\frac{d}{dt} \psi(t) = \left(H_{01} + \lambda H_2 \xi(t)\right)
\psi(t),
\end{eqnarray*}
where $\xi(t)=\frac{dW_t}{dt}$ is heuristically the time-derivative of the Brownian motion
$W_t$.
We have $\la \xi(t)\ra = 0$ and $\la \xi(t)\xi(t')\ra\, =
\delta(t-t')$ because the noise
should have zero mean and the noise at different times should be uncorrelated.
If we average over different realizations and define $\rho(t) = \la \rho_\xi \ra$
then $\rho (t)$ satisfies
\beq
\frac{d}{dt} \rho=-\frac{i}{\hbar}[H_{01},\rho]-\frac{\lambda^2}{2\hbar^2} [H_2,[H_2,\rho]].
\label{master}
\eeq
More details on the derivation can be found in the appendix.
 
We may consider the two effects together with the master equation
\begin{eqnarray}
\frac{d}{dt} \rho &=& -\frac{i}{\hbar} [H_0 + \beta H_1,\rho] - \frac{\lambda^2}{2 \hbar^2} [H_2,[H_2,\rho]], 
\end{eqnarray}
where $\beta$ is the amplitude of the systematic noise described by the
Hamiltonian $H_1$ and $\lambda$ is the strength of the amplitude noise.

In this paper, we assume that the errors affect the frequencies $\Omega_{R}$
and $\Omega_{I}$ but not the detuning $\Delta$, which, for an atom-laser 
realization of the two-level system is more easily controlled. 
For the systematic error we restrict ourselves to an error 
Hamiltonian of the form
\beqa
H_1 (t)= \frac{\hbar}{2} \left(\begin{array}{cc} 0 & \Omega_R (t) - i
  \Omega_I (t)\\
\Omega_R (t) + i \Omega_I (t) &  0
\end{array}\right)\, = H_0 (t)\Big|_{\Delta\equiv 0} .
\eeqa
For the noise error we restrict ourselves to independent amplitude-noise
in $\Omega_R$ as well as in $\Omega_I$ with the same intensity
$\lambda^2$, i.e. the final master equation is
\begin{eqnarray}
\fl \frac{d}{dt} \rho &=& -\frac{i}{\hbar} [H_0 + \beta H_1,\rho]
 -\frac{\lambda^2}{2 \hbar^2} \left([H_{2R},[H_{2R},\rho]] +
     [H_{2I},[H_{2I},\rho]]\right),
\label{masterfinal}
\end{eqnarray}
where
\begin{eqnarray*}
H_{2R} (t)= \frac{\hbar}{2} \left(\begin{array}{cc} 0 & \Omega_R (t)\\
\Omega_R (t) &  0
\end{array}\right), \,
H_{2I} (t)= \frac{\hbar}{2} \left(\begin{array}{cc} 0 & - i
  \Omega_I (t)\\
i \Omega_I (t) &  0
\end{array}\right)\,.
\end{eqnarray*}
A motivation for this modeling is that two different lasers may be used to 
implement the two parts of the Rabi frequency.

It is now convenient to represent the density matrix
$\rho(t)$ by the Bloch vector
\begin{eqnarray*}
\vec r (t) = \left(\begin{array}{c}
  \rho_{12}+\rho_{21}\\i(\rho_{12}-\rho_{21})\\\rho_{11}-\rho_{22}\end{array}\right),
\end{eqnarray*}
such that $\rho = \frac{1}{2} (1 + \vec r \cdot \vec\sigma)$ where
$\vec\sigma = (\sigma_1,\sigma_2,\sigma_3)$ are the Pauli matrices.
The Bloch equation corresponding to the master equation (\ref{masterfinal}) is
\begin{eqnarray}
\frac{d}{dt}\vec r = \left(\hat L_0 + \beta \hat L_1 - \lambda^2 \hat
L_2\right) \vec r,
\label{blochfinal}
\end{eqnarray}
where
\begin{eqnarray*}
\hat L_0 = \left(\begin{array}{ccc}
0 & \Delta (t) & \Omega_{I} (t) \\
-\Delta (t) & 0 & - \Omega_{R} (t)\\
-\Omega_{I} (t) & \Omega_{R} (t) & 0
\end{array}\right)\, , \,
\hat L_1 = \left(\begin{array}{ccc}
0 & 0 & \Omega_{I} (t) \\
0 & 0 & - \Omega_{R} (t)\\
-\Omega_{I} (t) & \Omega_{R} (t) & 0
\end{array}\right),
\end{eqnarray*} 
and
\begin{eqnarray*}
\hat L_2 = \frac{1}{2} \left(\begin{array}{ccc}
\Omega_{I}(t)^2 & 0 & 0 \\
0 & \Omega_{R}(t)^2 & 0\\
0 & 0 & \Omega_{R}(t)^2 + \Omega_{I} (t)^2
\end{array}\right).
\end{eqnarray*}
Note that the probability to be in the excited state at time $t$ is 
$P_2(t) = \frac{1}{2} (1-r_3(t))$.
In the following section we will first study the amplitude-noise errors only,
then in \sref{systematic_error} the
systematic errors and finally both together.

%
\section{Amplitude-noise errors}
\label{noise_error}
We assume that there is an amplitude-noise type or error 
affecting the Rabi frequencies and  no systematic errors ($\beta=0$). 
Let us define the noise sensitivity as
\begin{eqnarray*}
q_N := -\frac{1}{2} \left. \frac{\partial^2 P_2}{\partial \lambda^2}\right|_{\lambda=0}
= - \left.\frac{\partial P_2}{\partial (\lambda^2)}\right|_{\lambda=0},
\end{eqnarray*}
where $P_2$ is the probability to be in the excited state at final time $T$,
i.e. $P_2 \approx 1 - q_N \lambda^2$.
A smaller value of $q_N$ means less sensitivity with respect to amplitude-noise
errors, i.e. the scheme is more stable concerning this type of noise.
In general an analytic solution of the master equation (\ref{masterfinal}) or
the Bloch equation (\ref{blochfinal}) cannot be found. To 
calculate $q_N$ we do a perturbation approximation of the solution keeping
only terms up to $\lambda^2$ (with $\beta = 0$). In this manner we get
\begin{eqnarray*}
r_3 &=& (0,0,1) \vec r\\
&\approx& 
r_{0,3}(T) - \lambda^2 \int_0^T dt' (0,0,1) \tilde U_0(t,t') \hat L_2(t') \vec r_0(t')\\
&\approx& 
r_{0,3}(T) + \lambda^2 \int_0^T dt' (0,0,-1) \tilde U_0(t,t') \hat L_2(t') \vec r_0(t').
\end{eqnarray*}
where $\tilde U_0$ is the unperturbed time evolution operator for the Bloch vector. 
If the noiseless scheme works perfectly, i.e. $r_{0,3}(T) = -1$, then
\begin{eqnarray*}
P_2 = 1 - \frac{\lambda^2}{2} \int_0^T dt' \vec r_0 (t')^\cT \hat L_2(t') \vec r_0(t'),
\end{eqnarray*}
where $^\cT$ means the transpose operation
and the noise sensitivity becomes
\begin{eqnarray}
\fl q_N  &=& \frac{1}{2} \int_0^T dt \vec r_0 (t)^\cT \hat L_2(t) \vec r_0(t)\nonumber\\
\fl &=& \frac{1}{4} \int_0^T dt \Big[\Omega_{I}(t)^2 (r_{0,1}(t)^2 +
  r_{0,3}(t')^2)
 + \Omega_{R}(t')^2 (r_{0,2}(t')^2 +
  r_{0,3}(t')^2) \Big].
\label{eq_qn}
\end{eqnarray}

\subsection{Example: $\pi$ pulse with real Rabi frequency}
\label{noise_pi}
\begin{figure}[t]
\begin{center}
\includegraphics[width=0.45\linewidth]{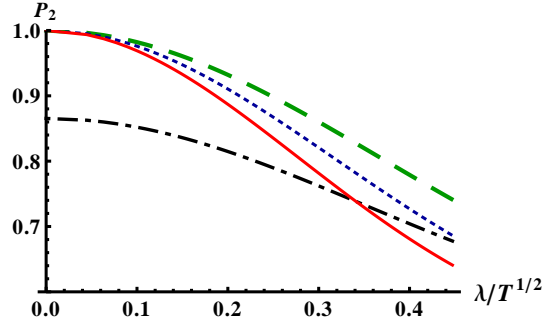}
\caption{(Color online) Probability $P_2$ to end in the excited state at time $T$
versus noise parameter $\lambda$.
Optimal protocol (green, dashed line), flat $\pi$-pulse with purely real
Rabi frequency (blue, dotted line),
pure adiabatic (black, dashed-dotted line), transitionless shortcut method (red, solid
line).
Additional parameter for adiabatic
protocol and transitionless protocol: $\Omega_0 T = 5.57/4.3\,\pi$, $\delta_0
T = (5.57/4.3)^2\pi$.}
\label{noise_ex1}
\end{center}
\end{figure}

As a first simple example of a population-inversion protocol we look at a $\pi$ pulse with a real Rabi
frequency, i.e. we set $\Delta = 0$, $\Omega_I = 0$ and $\int_{0}^{T} \Omega_R (t) dt= \pi$.
In this case the master equation resp. the Bloch equation 
for amplitude noise can be solved analytically.
The solutions of this equation with the initial conditions $\rho_{11}(0)=1$, $\rho_{12}(0)
=\rho_{21}(0)=\rho_{22}(0)=0$ resp. $\vec r(0) =
(0,0,1)^\cT$ at initial time $t=0$ are
\begin{eqnarray}
r_1 (t) &=& 0\nonumber\\
r_2 (t) &=& -e^{-\lambda^2 \int_{t_s}^t \Omega_a^2(t') dt'/2}
\sin\left(\int_{0}^t \Omega_a (t') dt'\right),
\nonumber\\
r_3 (t) &=&
e^{-\lambda^2 \int_{0}^T \Omega_a^2(t') dt'/2}
\cos\left(\int_{0}^t \Omega_a(t') dt' \right),
\label{eq_pi_real}
\end{eqnarray}
which yields
$$
P_2 = \frac{1}{2} - \frac{1}{2} r_3 (T) = \frac{1}{2}+\frac{1}{2}e^{-\lambda^2 \int_{0}^{T}
  \Omega_R^2(t) dt/2}.
$$
The noise sensitivity is now
\begin{eqnarray}
q_N = \frac{1}{4} \int_{0}^{T} \Omega_R^2(t') dt', 
\label{pi_qn}
\end{eqnarray}
which may be bounded as $\frac{\pi^2}{4T} \le q_N \le \frac{\pi}{4}
\max_{0 \le t \le T} |\Omega_R(t)|$, where the lower bound is derived using
Schwartz inequality,
i.e. $\fabsq{\int_0^T dt\, \Omega_R(t)} \le \int_{0}^{T} dt\, \Omega_R^2(t)$.
We can achieve the lower bound using a constant $\Omega_R = \pi/T$ (i.e. a flat $\pi$-pulse).
The excitation probability $P_2$ for this flat $\pi$-pulse is plotted in
\fref{noise_ex1} versus the noise intensity $\lambda$ (blue, dotted line).
The noise sensitivity is $q_N = \pi^2/4T \approx 2.467/T$.
The other lines in \fref{noise_ex1} correspond to different protocols, see below for more details. The important thing at this point is to note that 
the stability of a protocol is very well
quantified by $q_N$, which is the curvature at $\lambda=0$.

\subsection{Example of a transitionless protocol}
\label{trans_noise}
As another example, we will now look at the stability and noise sensitivity of
a transitionless shortcut protocol.
Our reference scheme is the finite-time sinusoidal model \cite{Lu,Xiao}
\begin{eqnarray}
\Omega_R (t) = \Omega_0 \sin\left(\frac{\pi t}{T}\right)\,,\quad
\Delta (t) = -\delta_0 \cos\left(\frac{\pi t}{T}\right),
\label{sinusoidal}
\end{eqnarray}
with $\Omega_I = 0$ and $0 \le t \le T$.
The excitation probability is also shown in
\fref{noise_ex1} (black, dashed-dotted line).  
The chosen intensities are not large
enough and the population inversion is not complete. 

\begin{figure}[t]
\begin{center}
\includegraphics[width=0.45\linewidth]{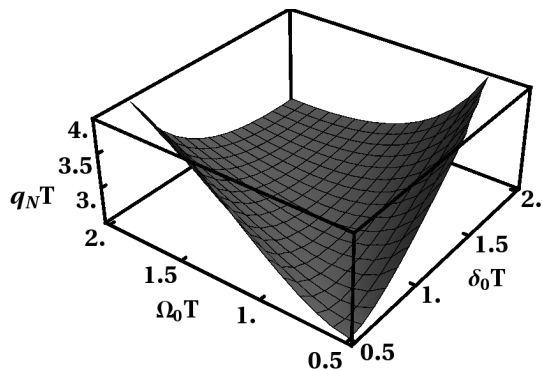}
\caption{Noise sensitivity $q_N$ versus $\Omega_0$
and $\delta_0$ for the transitionless protocol.}
\label{noise_ex2}
\end{center}
\end{figure}

\Fref{noise_ex1} shows the excitation probability
also for the transitionless shortcut based on this sinusoidal
model (red, solid line).
The noise sensitivity of this protocol is $q_N = 3.21/T$.
\Fref{noise_ex2} shows the noise sensitivity for the transitionless protocol
based on the sinusoidal model (\ref{sinusoidal})
for different values of $\delta_0$ and $\Omega_0$. The minimal noise
sensitivity in this figure is achieved for $\delta_0 = \Omega_0 = 0.5/T$ and
it has the value $q_N = 2.475/T$ which is very close to the noise sensitivity
of the flat $\pi$ pulse in the previous subsection.

%
%

\subsection{Optimal scheme}
We can also write the unperturbed Bloch vector in the  form
\begin{eqnarray}
\vec r_0(t) = \left(\begin{array}{c} \sin\Theta \cos\alpha\\
\sin\Theta\sin\alpha\\
\cos\Theta
\end{array}\right).
\label{bloch}
\end{eqnarray}
This Bloch vector corresponds to the pure state in (\ref{purestate}).
Therefore we get from the Bloch equation the same equations (\ref{schrpure}).
If the trajectory of the Bloch vector $\vec r(t)$ and $\Delta (t)$
is given, i.e. $\Theta, \alpha$ and $\Delta$ are given, then the
corresponding $\Omega_R$ and $\Omega_I$
can be calculated by (\ref{pot_R}) and (\ref{pot_I}).
Let $m(t) = \tan\Theta (\Delta + \dot\alpha)$.
Using (\ref{eq_qn}), (\ref{bloch}) and 
(\ref{pot_R})-(\ref{pot_D}), we get for the noise error sensitivity
\begin{eqnarray}
\fl q_N &=& \frac{1}{4} \int_0^T dt \Big[
(\cos^2\Theta + \cos^2\alpha\sin^2\Theta)(m\sin\alpha - \cos\alpha
    \dot\Theta)^2\\
\fl && + (\cos^2\Theta + \sin^2\alpha\sin^2\Theta)(m\cos\alpha + \sin\alpha
    \dot\Theta)^2\Big]\, \equiv \int_0^T dt L(m,\alpha,\Theta,\dot\Theta),
\end{eqnarray}
where $L$ is the Lagrange function for $q_N$.
We are looking for functions $m(t),\Theta(t),\alpha(t)$ which minimize this
functional. From the Euler-Lagrange formalism we get
\begin{eqnarray*}
0 = \frac{\partial L}{\partial m}
\Rightarrow  m= \frac{\dot\Theta \sin(4\alpha) \sin^2\Theta}
{4\cos^2\Theta + 2\sin^2(2\alpha)\sin^2\Theta}.
\end{eqnarray*}
Moreover
\begin{eqnarray*}
0 = \frac{\partial L}{\partial\alpha}
\Rightarrow \sin(4\alpha) = 0 \Rightarrow \alpha=n \pi/4 .
\end{eqnarray*}
From this it also follows that $m(t) = 0$.
Finally we have
\begin{eqnarray}
0 &=& \frac{\partial L}{\partial\Theta}- \frac{d}{dt} \frac{\partial
L}{\partial\dot\Theta}.
\label{eqtheta}
\end{eqnarray}
Let us now consider the cases $n$ odd and $n$ even separately.

\paragraph{Case $n$ even}
If $n$ is even, then (\ref{eqtheta}) simplifies to $\ddot\Theta = 0$.
Taking the boundary conditions $\Theta(0)=0, \Theta(T)=\pi$ into account, we
arrive at
\begin{eqnarray*}
\Theta(t) = \pi t/T.
\end{eqnarray*}
It follows that 
\begin{eqnarray*}
\Omega_R = -\sin\left(\frac{n\pi}{4}\right) \, \frac{\pi}{T}\,,\quad \Omega_I = \cos\left(\frac{n\pi}{4}\right) \,
\frac{\pi}{T}\,,\quad \Delta =0.
\end{eqnarray*}
Note that either $\Omega_R$ or $\Omega_I$ is zero, so these schemes are flat
$\pi$ pulses with a purely real or purely imaginary Rabi frequency.
(As an example, we get for $n=6$ a $\pi$ pulse with a flat, real Rabi
frequency $\Omega_R= \frac{\pi}{T}$ and $\Omega_I = 0$.)
For all these schemes an analytical solution of the master equation can be derived
similar to the one in subsection \ref{noise_pi} and the noise sensitivity of
all schemes is $q_N = \pi^2/(4T)$.

\begin{figure}[t]
\begin{center}
\includegraphics[width=0.45\linewidth]{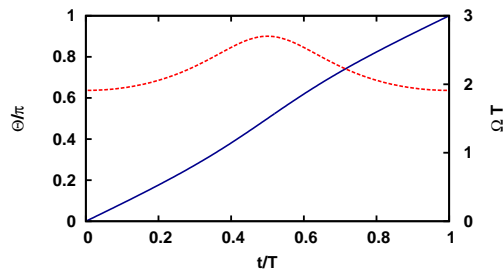}
\caption{(Color online) Protocol with minimal noise sensitivity $q_N$,
$\Theta(t)$ (blue, solid line; left axis), $\Omega(t)$ (red, dashed line; right axis).}
\label{optimal_noise}
\end{center}
\end{figure}

\paragraph{Case $n$ odd}
For $n$ odd we get
\begin{eqnarray}
(3 + \cos(2\Theta))\ddot\Theta = \sin(2\Theta) (\dot\Theta)^2
\label{eqx}
\end{eqnarray}
Then $\Omega_{R} = \pm \dot\Theta/\sqrt{2} = \pm \Omega_I$ and $\Delta(t)=0$. 
We first solve  (\ref{eqx}) for $\Theta$ numerically and then put the
solution in the expression for the noise sensitivity.
The numerically calculated $\Theta(t)$ can be seen
in \fref{optimal_noise} (solid line, left axis). The corresponding Rabi frequencies for
$n=7$ are $\Omega_R (t) = \Omega_I (t) = \Omega (t)$, where $\Omega$ is shown
also in \fref{optimal_noise} (dashed line, right axis). Note that for
other values of odd $n$ only the signs of $\Omega_R$ resp. $\Omega_I$
are switched. The noise sensitivity value is $q_N = 1.82424/T < \pi^2/(4T)$.
Therefore, for $n$ odd, smaller noise sensitivities can be achieved
than for $n$ even, so these protocols are
least sensitive to amplitude noise.
Finally, the optimal $\pi$ pulse is shown (green, dashed line) in
\fref{noise_ex1}, it has a noise sensitivity $q_N = 1.82424/T$.

Note that an approximate solution of (\ref{eqx}) is given by $\Theta(t) = \pi t/T -
\frac{1}{12}\sin(2\pi t/T)$, with a noise sensitivity of $q_N=1.82538/T$.

%
\section{Systematic errors}
\label{systematic_error}
In this section we shall only consider systematic errors, i.e. $\lambda=0$.
It is enough to work with pure states, instead of density
matrices, satisfying the Schr\"odinger equation
\begin{eqnarray*}
i\hbar\frac{d}{dt}|\psi(t)\ra = (H_0 (t) + \beta H_1) |\psi(t)\ra.
\end{eqnarray*}
We define the systematic error
sensitivity as
\begin{eqnarray*}
q_S := -\frac{1}{2}\left. \frac{\partial^2 P_2}{\partial \beta^2}
\right|_{\beta=0} = -\left. \frac{\partial P_2}{\partial (\beta^2)}
\right|_{\beta=0},
\end{eqnarray*}
where $P_2$ is the probability to be in the excited state at final time
$T$.

Using perturbation theory up to $O(\beta^2)$ we
get
\begin{eqnarray*}
\fl |\psi(T)\ra &=& |\psi_0(T)\ra - \frac{i}{\hbar}\beta \int_0^T dt \hat U_0(T,t) H_1(t') |\psi_0(t)\ra\\
\fl && - \frac{1}{\hbar^2}\beta^2 \int_0^T dt \int_0^{t'} dt'\, \hat U_0(T,t)
H_1(t') \hat U_0(t,t') H_1 L_1(t')  |\psi_0(t')\ra + ..., 
\end{eqnarray*}
where $|\psi_0(t)\ra$ is the unperturbed solution and $\hat U_0$ the
unperturbed time evolution operator.
We assume that the error-free ($\beta=0$) scheme works perfectly,
i.e. $|\psi_0(T)\ra= e^{i\mu} |2\ra$ with some real $\mu$.
Then, 
\begin{eqnarray*}
\fl P_2 &=& |\la 2|\psi(T)\ra|^2 = \la \psi (T) | \psi_0 (T) \ra\la \psi_0 (t) | \psi
(T)\ra
\approx 1 - \frac{\beta^2}{\hbar^2}
\left| \int_0^T dt \la \psi_\perp (t) | H_1 (t) | \psi_0 (t) \ra \right|^2
\end{eqnarray*}
because $\hat U_0(s,t) =  |\psi_0 (s) \ra \la \psi_0 (t) | + |\psi_\perp(s) \ra \la
\psi_\perp (t) |$,
where $\la \psi_\perp (t) | \psi_0 (t)\ra = 0$ for all times and $|\psi_\perp (t)\ra$
is also a solution of the Schr\"odinger equation, see (\ref{psip}).
From this we get the systematic-error sensitivity value
\begin{eqnarray*}
q_S &=& \frac{1}{\hbar^2}\left| \int_0^T dt \la \psi_\perp (t) | H_1 (t) | \psi_0 (t) \ra
\right|^2.
\end{eqnarray*}
\subsection{Example: $\pi$ pulse}
Let $\Delta = 0$, $\Omega (t)= |\Omega(t)| e^{i\alpha}$ and
$\int_0^T dt\, |\Omega(t)|=\pi$, which correspond to a $\pi$ pulse.
Then we get that $H_0 = H_1$ and an analytical solution exists,
\begin{eqnarray*}
P_2 &=& \frac{1}{2}-\frac{1}{2} \cos\left( (1+\beta) \int_{0}^{T} |\Omega(t')|
d t'\right) = \frac{1}{2}-\frac{1}{2} \cos\left( (1+\beta) \pi \right).
\end{eqnarray*}
It follows that $q_S = \pi^2/4$ independently of the time duration $T$.

\begin{figure}[t]
\begin{center}
\includegraphics[width=0.45\linewidth]{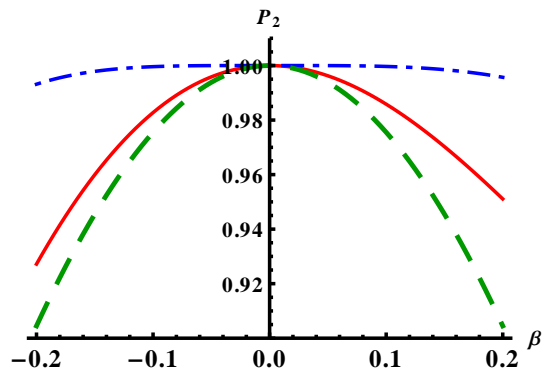}
\caption{(Color online)
Excitation probability $P_2$ versus systematic-error
  parameter $\beta$: protocol with zero systematic-error sensitivity (blue, dashed-dotted line),
transitionless protocol (red, solid line), $\pi$ pulse with minimal noise
sensitivity (green, dashed line).}
\label{sysex}
\end{center}
\end{figure}

The excitation probability versus
systematic noise $\beta$ is shown in \fref{sysex}.
As an example we are looking at the $\pi$ pulse which was optimal for
amplitude-noise error in the previous section (green, dashed line).
It has the systematic-error sensitivity $q_S = \frac{\pi^2}{4}\approx2.47$
which is equal for all $\pi$ pulses. Even if the protocol is 
maximally robust concerning amplitude-noise, it is very sensitive to systematic
errors.
%
%

\subsection{Example of a transitionless shortcut}
We again look at the example of a transitionless shortcut based on the sinusoidal model
which was
examined in subsection \ref{trans_noise}.
The excitation probability versus systematic noise $\beta$ is shown in
\fref{sysex} (red solid line).
The transitionless shortcut based on the sinusoidal model
is more stable concerning systematic errors that any $\pi$ pulse.

\begin{figure}[t]
\begin{center}
\includegraphics[width=0.45\linewidth]{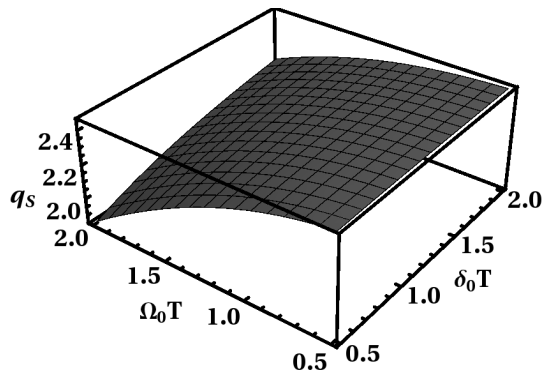}
\caption{Systematic-error sensitivity $q_S$ versus $\Omega_0$
and $\delta_0$ for the transitionless protocol;
the systematic-error sensitivity for a $\pi$ pulse corresponds to the upper $x$-$y$ plane.}
\label{sys_ex2}
\end{center}
\end{figure}

\Fref{sys_ex2} shows the systematic-error sensitivity for the transitionless-based protocol
for different values of $\delta_0$ and $\Omega_0$. Again, the protocol takes as a 
reference the sinusoidal model (\ref{sinusoidal}). Note that the
systematic-error sensitivity $q_S$ for any $\pi$ pulse corresponds to the upper $x$-$y$ plane in
this figure. This means that for all parameters shown the transitionless
shortcut is less sensitive (i.e. more stable) concerning systematic error than
any $\pi$ pulse.

%

\subsection{Optimal scheme}
To find an optimal scheme we shall use the invariant based technique. 
The pure state $|\psi(t)\ra$ can be parameterized as in  (\ref{purestate}).
The boundary values should be $\Theta(0)=0$ and $\Theta(T)=\pi$.
We get for the functions in the Hamiltonian leading to this solution 
\begin{eqnarray*}
\Omega_R &=& \cos\alpha\sin\Theta \;\dot\gamma -
\sin\alpha\;\dot\Theta\\
\Omega_I &=& \sin\alpha\sin\Theta\;\dot\gamma +
\cos\alpha\;\dot\Theta\\
\Delta &=& -\cos\Theta \;\dot\gamma - \dot\alpha
\end{eqnarray*}
Note that, contrary to \sref{noise_error}, it is now more convenient to take  $\gamma(t)$ as a given function
instead of $\Delta(t)$.
A solution which is orthogonal to (\ref{purestate}), i.e. $\la
\psi_\perp|\psi\ra = 0$ for all times, is given by \eref{psip}.
The expression for the systematic error sensitivity is now
\begin{eqnarray*}
q_S &=& \left| \int_0^T dt \la \Psi_\perp (t) | H_1 (t)/\hbar | \psi (t) \ra \right|^2\\
&=& \frac{1}{4} \left|\int_0^T dt \left[ -i e^{-i\gamma} \dot\gamma \;
  \cos\Theta\sin\Theta + e^{-i\gamma} \dot\Theta \right]\right|^2\\
&=&  \frac{1}{4} \left|\int_0^T dt \left[ e^{-i\gamma} \;
  \frac{d}{dt}(\cos\Theta\sin\Theta) + e^{-i\gamma} \dot\Theta
  \right]\right|^2,
\end{eqnarray*}
where we have applied partial integration in the last step taking into
account the boundary values $\Theta(0)=0$ and $\Theta(T)=\pi$.
The expression can be further simplified and we get finally
\begin{eqnarray*}
q_S &=& \left|\int_0^T dt e^{-i\gamma}\dot\Theta \sin^2\Theta\right|^2.
\end{eqnarray*}

In the special case when $\gamma(t)$ is constant in time, we get
$q_S = \frac{\pi^2}{4}$ independently of $\Theta(t)$.
With the choice $\alpha(t)$ constant we recover the $\pi$ pulse.

\begin{figure}[t]
\begin{center}
\includegraphics[width=0.45\linewidth]{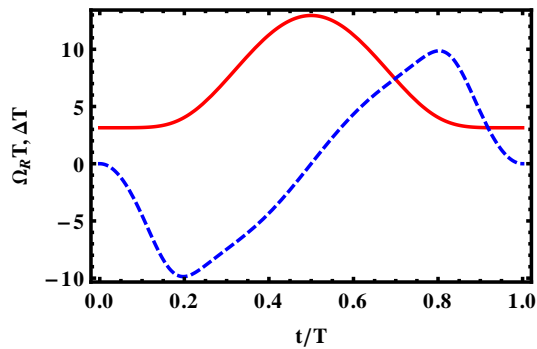}
\caption{(Color online) Rabi frequency $\Omega_R$ (red, solid line) and
  detuning $\Delta$ (blue, dashed line) for a protocol which has zero
  systematic-error sensitivity.}
\label{optsys}
\end{center}
\end{figure}

The minimum of $q_S$ is clearly achieved if $q_S=0$.
In the following we will show that there are protocols which fulfill this
condition, i.e. protocols maximally stable with respect to systematic errors.
We will give an example class which fulfills $q_S=0$.
Let
\begin{eqnarray*}
\gamma(t) = n \left(2 \Theta - \sin(2\Theta)\right).
\end{eqnarray*}
For this choice of $\gamma$ we get
\begin{eqnarray*}
q_S = \frac{\sin^2\left(n\pi\right)}{4n^2}.
\end{eqnarray*}
So for $n=1,2,3,...$ we get protocols fulfilling $q_S = 0$.
Note that in the limit of $n\to 0$ (i.e. $\gamma \to 0$), we get $q_S \to
\frac{\pi^2}{4}$, which is consistent with the previous paragraph.
The functions in the Hamiltonian in this case are
\begin{eqnarray*}
\Omega_R &=& \left(4n \cos\alpha\sin^3\Theta -
\sin\alpha\right)\;\dot\Theta ,\\
\Omega_I &=& \left(4n\sin\alpha\sin^3\Theta +
\cos\alpha\right)\;\dot\Theta ,\\
\Delta &=& -4n\cos\Theta\sin^2\Theta - \dot\alpha .
\end{eqnarray*}
Note that this class of protocols might not be the only ones fulfilling $q_S = 0$.
There is still some freedom left. For example, one could in addition
require that $\Delta=0$ or $\Omega_I=0$. In the following we will look at the
second condition, i.e. $\Omega_I=0$ for all $t$. This leads to 
$\alpha(t) = -\mbox{arccot}\left(4n\sin^3\Theta\right)$. In addition, there is the freedom
to chose $\Theta(t)$ with $\Theta(0)=0,\Theta(t_f)=\pi$.
For $n=1$ and $\Theta(t) = \pi t/t_f$, the resulting Rabi frequency and the
detuning are shown in \fref{optsys}.

%
%
\section{Systematic and amplitude-noise errors}
\label{noise_systematic}
\begin{figure}[t]
\begin{center}
(a)\includegraphics[width=0.5\linewidth]{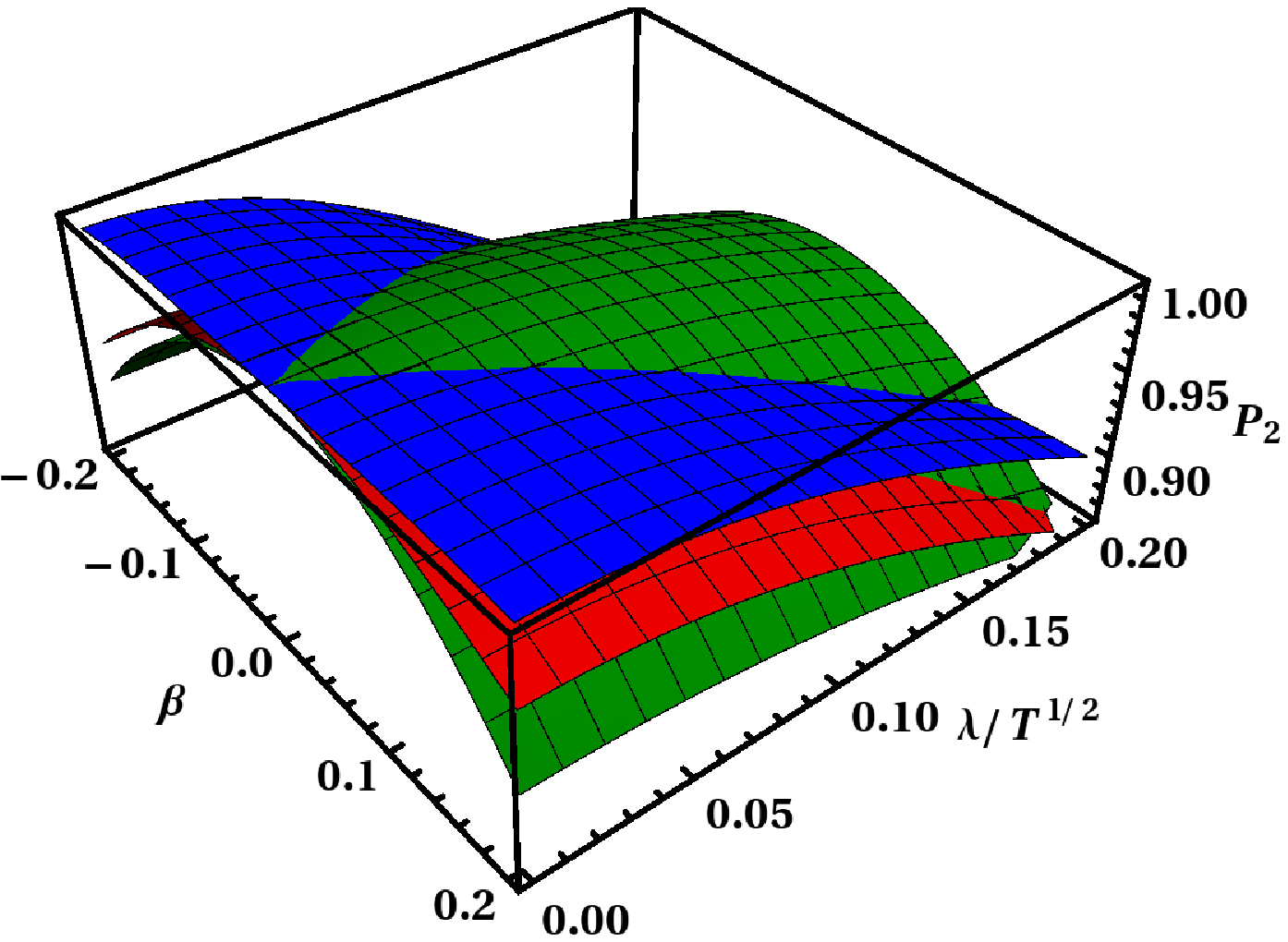}
\\[0.5cm]

(b)\includegraphics[width=0.29\linewidth]{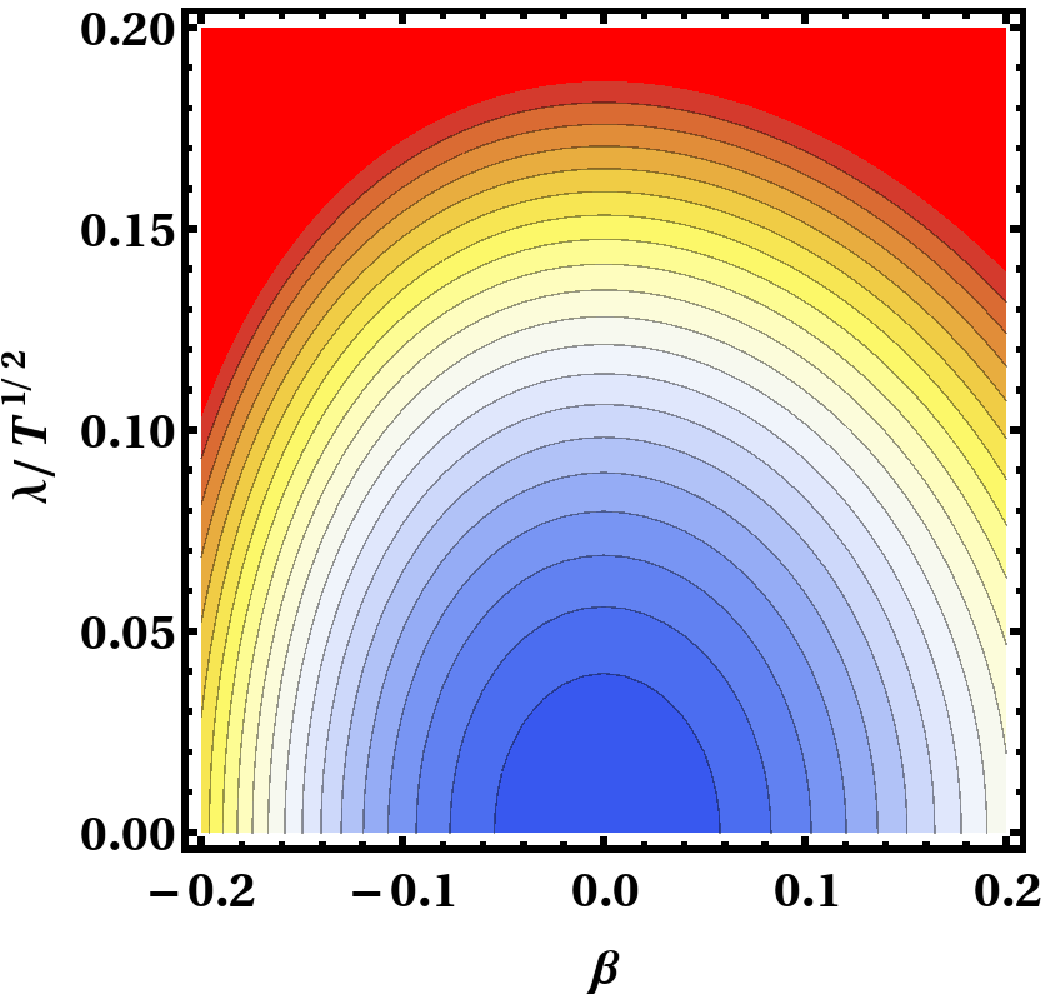}\,
(c)\includegraphics[width=0.29\linewidth]{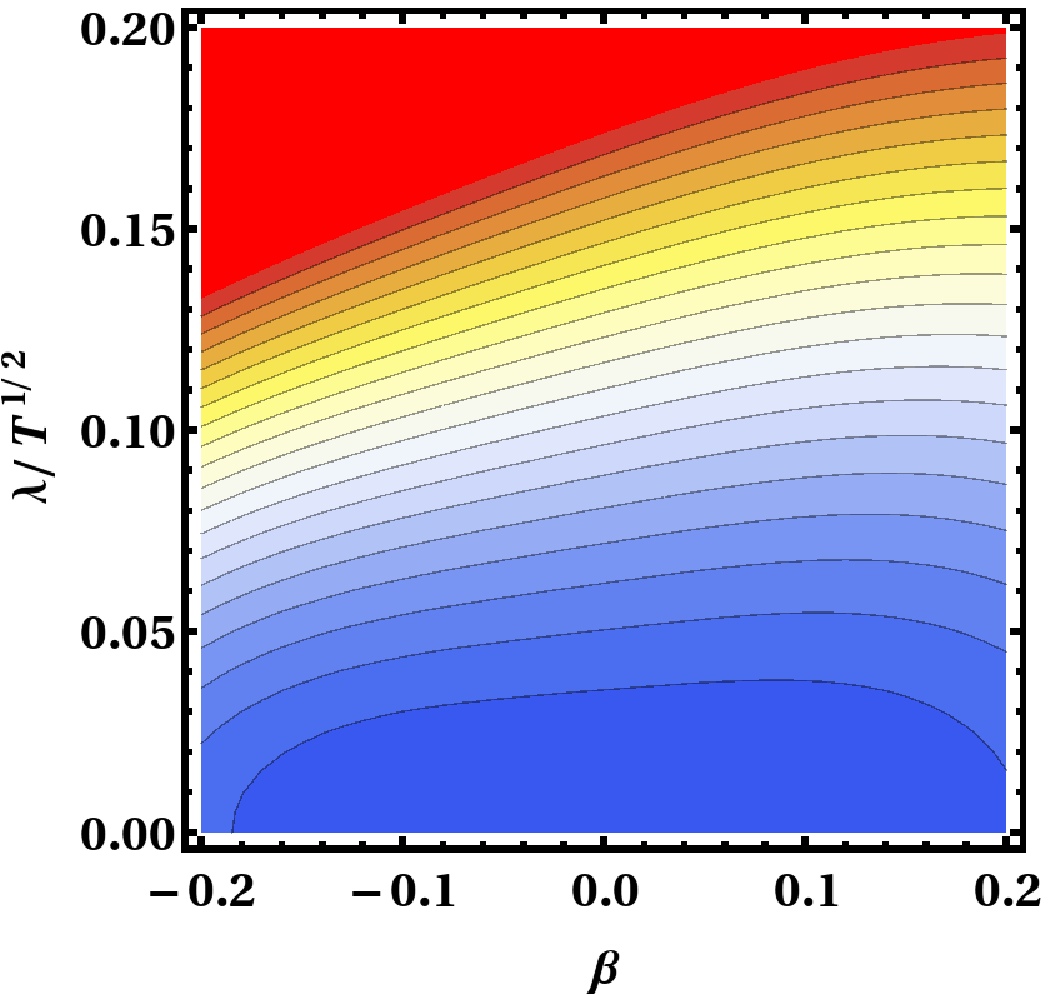}\,
(d)\includegraphics[width=0.29\linewidth]{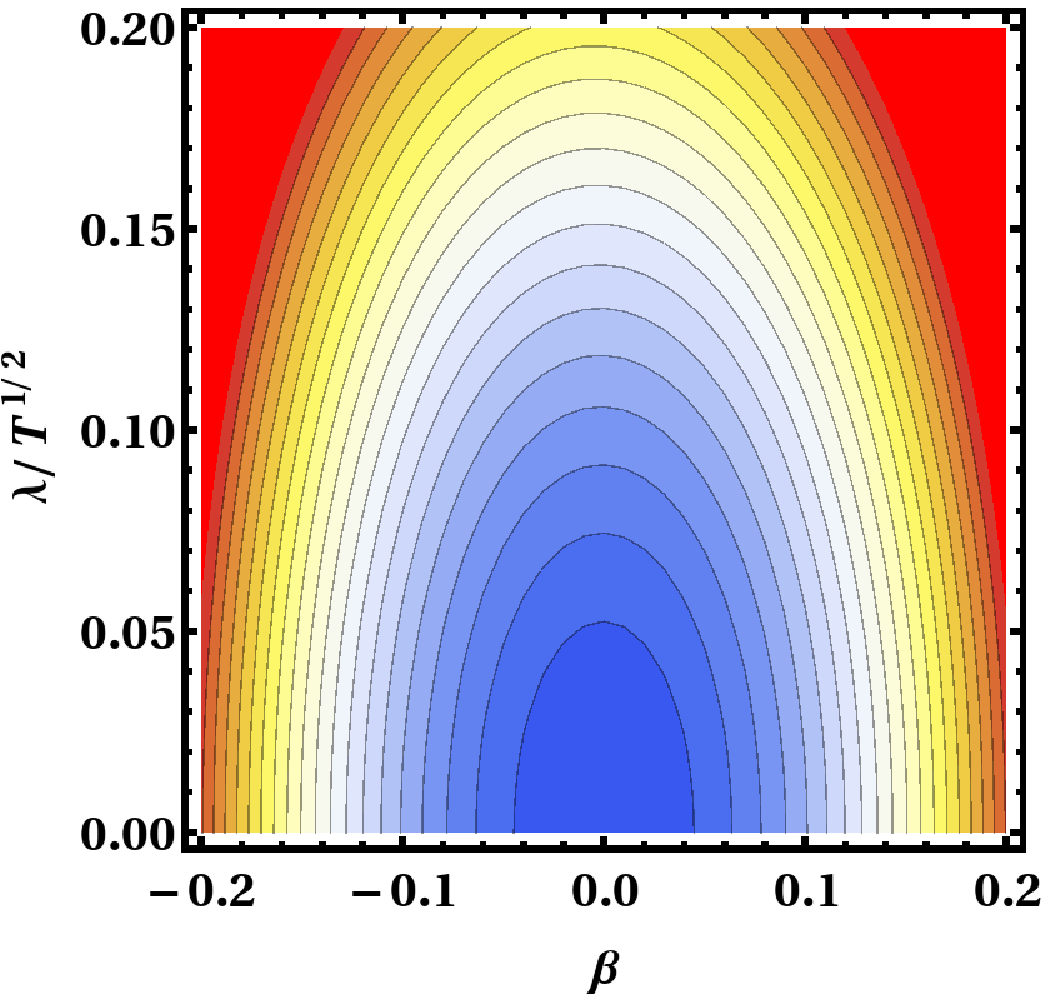}

\caption{(Color online) Probability $P_2$ versus noise error and systematic
  error parameter; (a) transitionless protocol (red),
optimal systematic stability protocol (blue), optimal noise protocol (green);
same result as contour plots:
(b) transitionless protocol, (c) optimal systematic stability protocol,
(d) optimal noise stability protocol.}
\label{general_ex}
\end{center}
\end{figure}

Finally, we will consider both types of errors together. 
Optimal schemes in this case would depend on the ratio between amplitude-noise
error and systematic error in the experiment. We will just examine numerically the
behavior of some protocols with respect to amplitude-noise and systematic error. 
Specifically we compare the minimal noise error protocol, the
minimal systematic error protocol, and the example of a transitionless
shortcut studied before, see \fref{general_ex}.
The figure shows that the different optimal schemes perform better than the other one 
depending on the dominance of one or the other type of error. 
%
%

Summarizing, in this paper we have examined the stability of different fast protocols
for exciting a two-level system with respect to 
amplitude-noise error and systematic errors.
First we have looked at the noise error alone and we have introduced a noise
sensitivity. We have shown that a special type of $\pi$
pulse is the optimal protocol with minimal noise sensitivity.
Then we have looked at the systematic error alone and we have introduced a
systematic error sensitivity. We have shown that there are protocols for which
this sensitivity is exactly zero.
Finally, we have looked at the general case with noise and systematic errors
together.
 
Future work may involve extending the present results to different types of noise and perturbations. The existence of a set of optimal solutions for systematic errors 
also opens the way to further optimization with respect to other variables of physical
interest.


\begin{appendix}
\section{Derivation of the Master equation for Amplitude-Noise Error}
The evolution of the quantum state with amplitude noise can only be described
by a master equation \cite{carmichael}. We assume that the Hamiltonian has a
deterministic part $H_{01}$ and a stochastic part containing $\lambda H_2$.
We need a mapping from a fixed time to another infinitesimally close, so our
starting point will be
\begin{equation}
|\psi_{t+dt}\rangle=e^{-i(H_{01} dt+\lambda H_2 dW_t)} |\psi_{t}\rangle
\label{eq:1}
\end{equation}
where $dt$ is the infinitesimal time step and $dW_t$ the corresponding noise
increment in the Ito sense. The properties of such noise are: 
$\la dW\ra=0, \la dW^2\ra=dt$. If we expand in Taylor series (\ref{eq:1}) and keep terms up to first
order in $dt$ and $dW$ (using the Ito calculus rules) we arrive at the
following Stochastic Schr\"odinger equation (SSE)
\beq
|d\psi\rangle = -\frac{i}{\hbar} H_{01} dt |\psi\rangle-\frac{\lambda^2}{2\hbar^2} H_2^2 dt |\psi\rangle -\frac{i\lambda}{\hbar} H_2 dW_t  |\psi\rangle.
\eeq
The master equation derived from this SSE is then (\ref{master}).

An equivalent approach in the Stratonovich sense is to start from
\begin{eqnarray*}
i\hbar\frac{d}{dt} \psi(t) = \left(H_{01} + \lambda H_2 \underbrace{\frac{dW_t}{dt}}_{\xi(t)}\right)
\psi(t),
\end{eqnarray*}
where $\xi(t)$ is heuristically the time-derivative of the Brownian motion
$W_t$.
We have $\la \xi(t)\ra = 0$ and $\la \xi(t)\xi(t')\ra\, =
\delta(t-t')$
because the noise
should have zero mean and the noise at different times should be uncorrelated.
If we average over different realizations and define $\rho(t) = \la \rho_\xi\ra$ then $\rho (t)$ is fulfilling  (\ref{master}).  
To show this we define $\rho_\xi (t) = |\psi_\xi(t)\ra \la\psi_\xi(t)|$.
We start from the dynamical equation for $\rho_{\xi}$, namely
\begin{equation}
\frac{d}{dt} \rho_{\xi}=-\frac{i}{\hbar} [H_{01},\rho_{\xi}]-\frac{i\lambda}{\hbar} [H_2,\xi \rho_{\xi}],
\label{eq:d1}
\end{equation}
that after averaging over the noise becomes
\begin{equation}
\frac{d}{dt} \rho=-\frac{i}{\hbar} [H_{01},\rho]-\frac{i\lambda}{\hbar} [H_2,\la \xi \rho_{\xi} \ra].
\label{eq:d2}
\end{equation}
Novikov's theorem applied to white noise takes the form
\begin{eqnarray*}
\la\xi(t)F[\xi]\ra = \frac{1}{2} \left<\frac{\delta F}{\delta\xi(s)}\right>_{s=t}.
\end{eqnarray*}
Using it we get
\begin{eqnarray}
	\la \xi \rho_{\xi} \ra=-\frac{i\lambda}{2\hbar} [H_2,\rho],
\end{eqnarray}
which leads to (\ref{master}).
\end{appendix}

\ack
We acknowledge funding by Projects No. GIU07/40, No. FIS2009-12773-C02-01,
No. FIS2010-19998, NSFC No. 61176118, and the UPV/EHU
under program UFI 11/55.\\

\end{document}